\documentclass[fleqn,12pt,twoside]{article}
\usepackage{espcrc1}

\usepackage{graphicx}
\usepackage{floatflt}

\newcommand{\AmS}{{\protect\the\textfont2
  A\kern-.1667em\lower.5ex\hbox{M}\kern-.125emS}}

\title{High $p_{T}$ Inclusive Charged Hadron Spectra from Au+Au Collisions 
at $\sqrt{s_{NN}}$ = 200 GeV}

\author{Jennifer L. Klay\address[LBNL]{Lawrence Berkeley National Laboratory,\\
        One Cyclotron Road, MS-70R0319, Berkeley, California, 94720-8169 U.S.A.}
        for the STAR Collaboration\thanks{For the full author list and acknowledgments, see Appendix ``Collaborations'' of this volume.}}
       
\begin{document}

\maketitle

\begin{abstract}
The STAR Collaboration presents new measurements of inclusive charged hadron
distributions for $p_{T} <$ 12 GeV/c from Au+Au collisions at $\sqrt{s_{NN}}$ = 200
GeV.  Charged hadron suppression at high $p_{T}$ is similar in
shape and magnitude at all centralities to that observed previously at $\sqrt{s_{NN}}$ = 130 GeV for $p_{T} <$ 6 GeV/c.
The ratio of spectra from central and peripheral Au+Au collisions shows 
that hadron suppression is approximately constant within  6 $< p_{T} <$ 12 GeV/c.  The ratios of charged hadron spectra at the two beam 
energies show a 15-20\% increase in yield at low $p_{T}$.  At high $p_{T}$, the ratios show a larger increase that 
agrees well with pQCD calculations of the $\sqrt{s_{NN}}$ dependence of particle production in Au+Au collisions. 
\end{abstract}

\section{Introduction}

The suppression of high $p_{T}$ hadron production at $\sqrt{s_{NN}}$ = 130 GeV observed by both the PHENIX
and STAR collaborations provided the first experimental signatures potentially related to partonic energy loss in the dense 
nuclear medium produced in RHIC collisions\cite{PHENIX,STAR}.  Theoretical calculations of the $p_{T}$ dependence of the nuclear 
modification factor, $R_{AA}$, suggest that an interplay of nuclear gluon shadowing, Cronin enhancement and partonic energy loss in 
a dense medium will cause the suppression to saturate at a finite value\cite{Vitev}.  Above a certain $p_{T}$ threshold, $R_{AA}$ may 
increase slowly with $p_{T}$ toward the binary scaling limit predicted by perturbative QCD\cite{XNWang}.  The new STAR data on 
charged hadron spectra in Au+Au collisions at $\sqrt{s_{NN}}$ = 200 GeV, which extend to $p_{T}$ = 12 GeV/c,
address these predictions and present an important constraint on theoretical descriptions of high $p_{T}$ phenomena in nuclear collisions.

\section{Data Collection and Analysis}

The data at $\sqrt{s_{NN}}$ = 200 GeV were taken during the second RHIC run in 2001 using the STAR detector.  
The minimum bias dataset, corrected for vertex finding efficiency, contains 97 $\pm$ 3\% of the 
total geometric cross-section for Au+Au collisions, assumed to be 7.2 
barns.  For this 
analysis, 2.7M minimum bias and 2.1M central triggered Au+Au events were used.  Event 
centrality classes corresponding to percentiles of the total cross-section 
were 
determined using the charged particle multiplicity within $|\eta| <$ 0.5.

Charged particles are detected in the TPC and their momentum is measured 
through their curvature in the 0.5 Tesla magnetic field.  The high $p_{T}$ 
analysis is restricted to long tracks ( $\ge$ 20 measured track points) 
with $|\eta| <$ 0.5 whose projected distance of closest approach to the 
primary event vertex falls within 1~cm of the measured vertex.

\begin{floatingfigure}{80mm}
\includegraphics*[width=75mm]{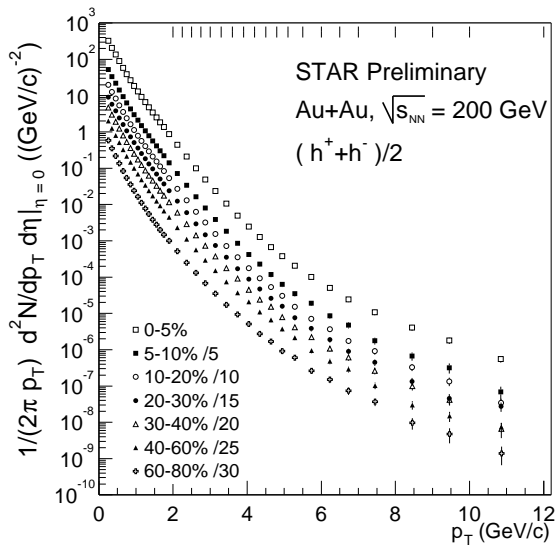}
\vspace{-2mm}
\caption{Charged hadron spectra from Au+Au collisions at $\sqrt{s_{NN}}$ = 200 GeV.}
\vspace{1mm}
\label{fig:Spectra}
\end{floatingfigure}

Tracking performance at high $p_{T}$ was assessed by embedding simulated tracks into real data events.  For the 0-5\% 
most central events, the efficiency and acceptance for tracks having $p_{T} >$ 3 GeV/c and $|\eta| <$ 0.5 are uniform 
as a function of $p_{T}$ at approximately 80\%, with a small dependence on 
pseudorapidity of $\pm$ 3\%.  For the 40-80\% most peripheral events, the efficiency is higher due to lower track density, and is 90\% 
above $p_{T}$ = 3 GeV/c and independent of $p_{T}$ and $\eta$.  Momentum 
resolution in the 200 GeV dataset is a factor of 3 better than 
that in the 130 GeV dataset\cite{STAR}.  

The dominant backgrounds in the high $p_{T}$ spectra are from weak decays of $\Lambda$ and $K^{0}_{s}$ particles 
and apparent high $p_{T}$ tracks from $\overline{p}$ and $\overline{n}$ annihilation in detector material.  Background 
corrections are estimated using HIJING events in a GEANT simulation of the detector, with the parent $\Lambda$, 
$K^{0}_{s}$ and $\overline{p}$ distributions from HIJING scaled to match STAR measured yields and slopes at 
$\sqrt{s_{NN}}$ = 130 GeV, corrected for the $\sqrt{s_{NN}}$ dependence of particle production.  The backgrounds are 
estimated to be less than 10\% for $p_{T} >$ 3 GeV/c.  However, due to the uncertainty in the yields of $\Lambda$s 
and $K^{0}_{s}$s at high $p_{T}$, the background corrections above $p_{T}$ = 3 GeV/c are assigned 
100\% uncertainty.  

The systematic uncertainties assigned to the spectra are determined by combining the uncertainties from the momentum resolution 
correction, the background subtraction and variations in the spectra for different sets of quality cuts and corrections.

\vspace{-1mm}
\section{Results}

The charged hadron $p_{T}$ spectra for $p_{T} <$ 12 GeV/c are shown in Fig.~\ref{fig:Spectra}, with the non-central bins scaled for 
clarity.  The limits of the bins in $p_{T}$ are indicated at the top of 
the figure by the hash marks, with the highest $p_{T}$ bin having the 
range 10 $< p_{T} <$ 12 GeV/c.   

\begin{floatingfigure}{80mm}
\includegraphics*[width=75mm]{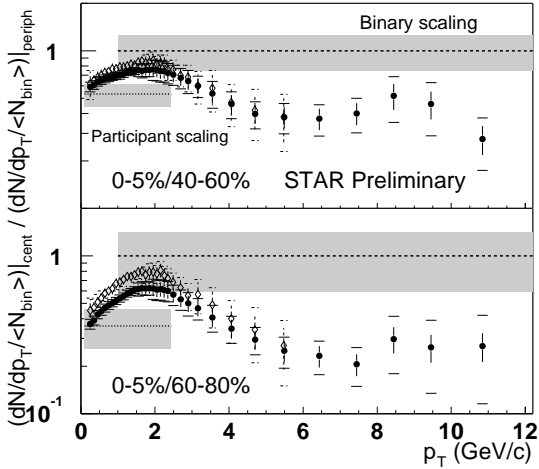}
\vspace{-2mm}
\caption{Central/peripheral ratio: upper panel is 0-5\%/40-60\%, lower panel is 0-5\%/60-80\%.}
\vspace{0.75mm}
\label{fig:CentPeriphRatio}
\end{floatingfigure}

To explore the pattern of hadron suppression to the highest $p_{T}$, Fig.~\ref{fig:CentPeriphRatio} compares the binary-collision 
scaled STAR central 0-5\% spectrum to the 40-60\% (top panel) and the 60-80\% (lower panel) spectra.  The preliminary 200 GeV data are shown as black circles, while the STAR results at $\sqrt{s_{NN}}$ 
= 130 GeV\cite{STAR} are shown as open diamonds.  The vertical error bars represent statistical and systematic uncertainties in the 
central 0-5\% spectra, while the caps show the quadrature sum of these uncertainties with the uncertainties associated with the 
peripheral reference spectrum.   The dashed lines indicate the mean number of binary collisions, $\langle N_{binary} \rangle$, and the mean number of 
participant nucleons, $\langle N_{part} \rangle$, obtained from a Monte Carlo Glauber model of the nuclear overlap, with their uncertainties shown as the gray 
boxes.  The suppression pattern is similar at the two beam energies, showing an increasing suppression as a 
function of $p_{T}$ for $p_{T}$ = 2 - 6 GeV/c.  The 200 GeV data, which extend to $p_{T}$ = 12 GeV/c, show that the suppression for 
central collisions saturates and is approximately constant above $p_{T}$ = 6 GeV/c.

The nuclear modification factor $R_{AA} = \frac{ dN/dp_{T}d\eta (AA)}{T_{AA} d\sigma/dp_{T}d\eta (NN)}$, where $T_{AA}$ is the nuclear overlap integral, 
depends on the knowledge of the nucleon-nucleon (NN) reference spectrum.  A power law parameterization ($d\sigma/dp_{T}d\eta = A (1+p_{T}/p_{0})^{-n}$) of the 
UA1 measured charged particle spectrum from $\overline{p}+p$ collisions at $\sqrt{s_{NN}}$ = 200 GeV\cite{UA1} is used to construct $R_{AA}$.  
The parameterization was corrected for the larger UA1 $|\eta|$ acceptance ($|\eta| <$ 2.5) as in\cite{STAR}.  Since power law fits cannot be 
meaningfully extrapolated beyond the range constrained by data, we restrict $R_{AA}$ analysis to $p_{T} <$ 6 GeV/c.  

\begin{figure}[htb]
\begin{minipage}{78mm}
\includegraphics*[width=75mm]{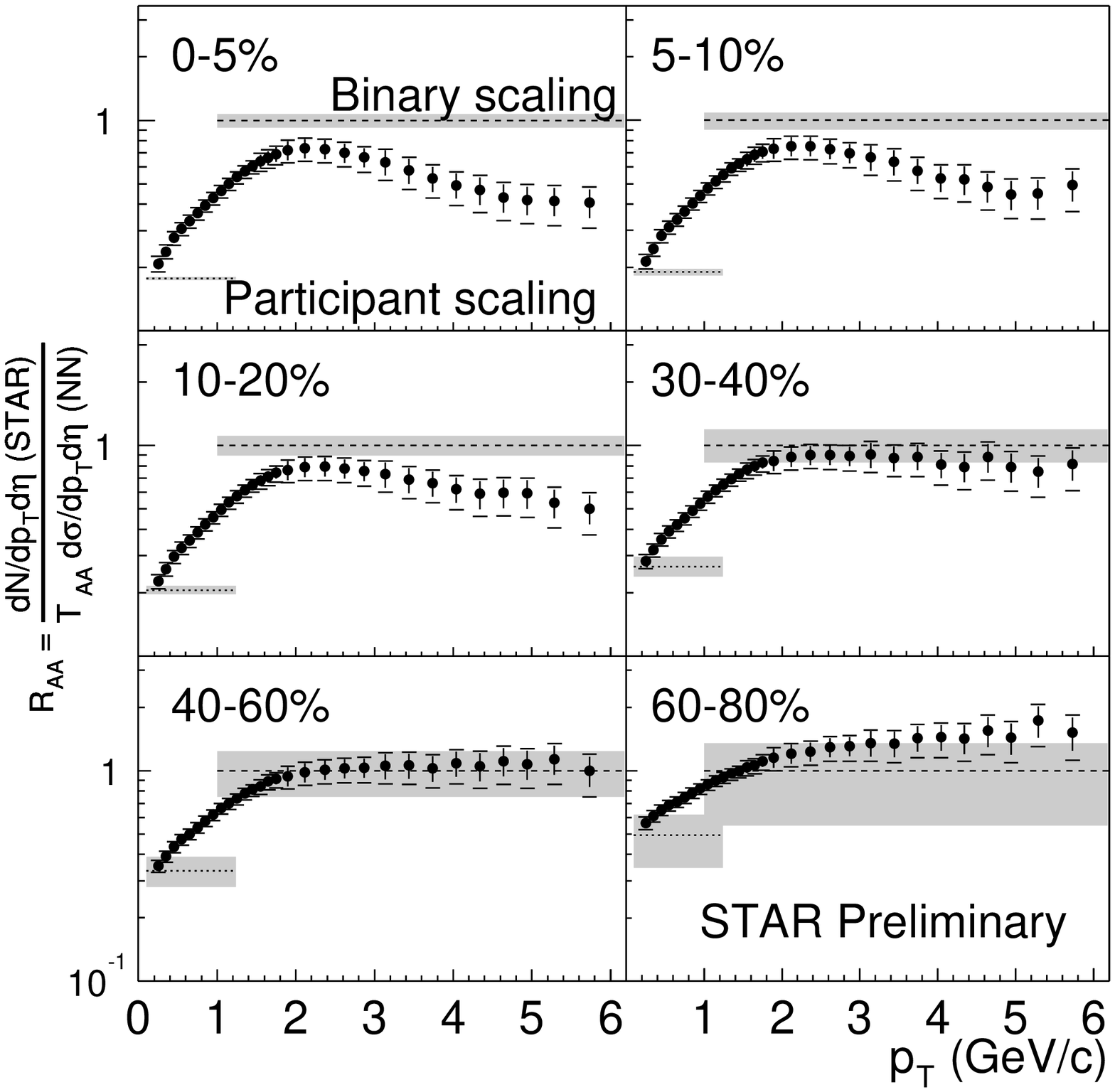}
\vspace{-5mm}
\caption{The nuclear modification factor, $R_{AA}$, as a function of $p_{T}$ for six centrality classes.  The NN reference spectrum 
is the same in all panels.}
\label{fig:AllRAA}
\end{minipage}
\hspace{\fill}
\begin{minipage}{78mm}
\includegraphics*[width=75mm]{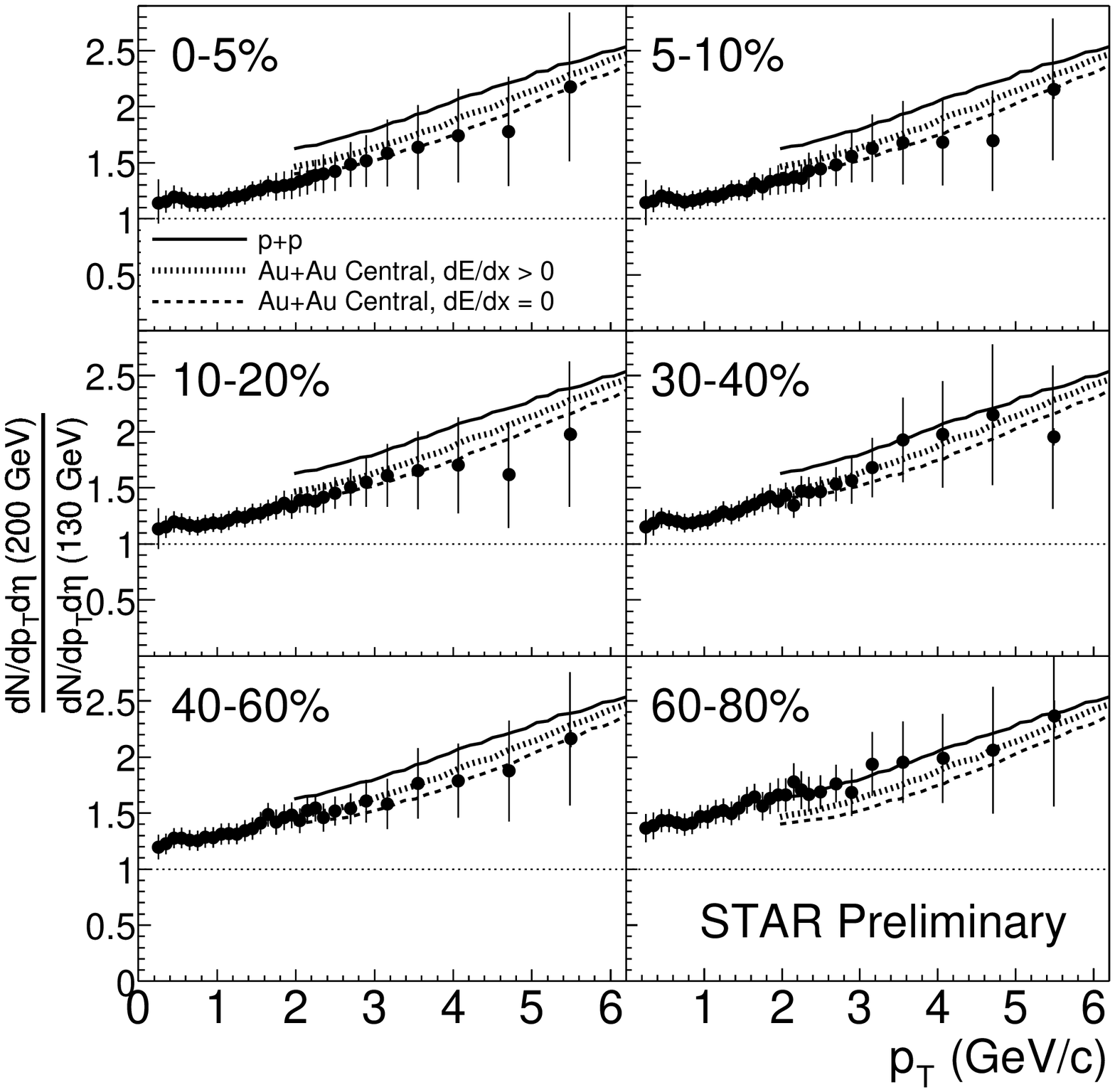}
\vspace{-5mm}
\caption{Ratios of charged hadron spectra from Au+Au collisions at $\sqrt{s_{NN}}$ = 200 and 130 GeV.  The curves (same in all panels) are pQCD 
calculations described in the text.}
\label{fig:RootSRatio}
\end{minipage}
\end{figure}

Fig.~\ref{fig:AllRAA} shows the centrality dependence of $R_{AA}$ at $\sqrt{s_{NN}}$ = 200 GeV.
The error bars are similar to those in Fig. \ref{fig:CentPeriphRatio}.  Plotting this ratio on a log scale separates the evolution of the 
shape of the measured suppression from the uncertainties in the geometric scaling and the reference spectrum.  The shape of the suppression is seen to 
evolve smoothly as a function of centrality, with no evidence of a threshold for the onset of suppression\cite{XNWang2}.  The magnitude and evolution 
of $R_{AA}$ at $\sqrt{s_{NN}}$ = 200 GeV is consistent with that observed in $\sqrt{s_{NN}}$ = 130 GeV Au+Au collisions~\cite{STAR}.

The $\sqrt{s_{NN}}$ dependence of charged particle production at midrapidity is shown in Fig.~\ref{fig:RootSRatio}.  The error bars 
represent the combination of statistical uncertainties in the numerator and the total uncertainty (statistical + systematic) in the 
denominator.  There is an approximately 15-20\% increase in the particle yield at low $p_{T}$ in the 200 GeV data 
compared to the 130 GeV data\cite{PHOBOS,Gene}.  The curves (the same in all panels) are pQCD 
calculations for p+p collisions (solid line) and central Au+Au collisions with (dotted line) and without (dashed line) quenching\cite{XNWang3}.
The high $p_{T}$ behaviour is consistent with the pQCD expectation.  The combination of these results with the 
$\sqrt{s_{NN}}$ dependence of the high $p_{T}$ charged hadron azimuthal anisotropy\cite{Kirill} and the high $p_{T}$ 
two-particle correlations\cite{Dave} observed by STAR, suggests a picture of strong absorption in the bulk and surface 
emission of hard scattering products.

\vspace{-2mm}
\section{Conclusions}

STAR has measured charged particle production within 0.2 $< p_{T} <$ 12 GeV/c as a function of centrality in Au+Au collisions at 
$\sqrt{s_{NN}}$ = 200 GeV.  The suppression observed at $\sqrt{s_{NN}}$ = 200 GeV is similar to that observed at 130 GeV.  With the extension 
of the spectra to much higher $p_{T}$, the binary-scaled central/peripheral ratios show that the charged hadron 
suppression is approximately constant 
above $p_{T}$ = 6 GeV/c.  The ratios of the spectra at $\sqrt{s_{NN}}$ = 200 GeV to those at 130 GeV show that high $p_{T}$ particle 
production grows with $\sqrt{s_{NN}}$ for all centralities in Au+Au collisions similarly to pQCD expectations.  In order to disentangle the 
interplay between nuclear gluon shadowing, Cronin enhancement and possible partonic energy loss in a high density quark gluon 
plasma, which may all contribute to the observed suppression pattern, d+A collisions at RHIC are eagerly anticipated.

\end{document}